\documentclass[twocolumn,amsmath,amssymb,amsfonts,longbibliography]{revtex4-1}

\usepackage{graphicx}
\usepackage{dcolumn}
\usepackage{bm}
\usepackage{latexsym,color}

\usepackage{braket}

\usepackage{hyperref}
\usepackage{natbib}

\hypersetup{
  colorlinks,
  citecolor=blue,
  linkcolor=blue,
  urlcolor=blue}

\newcommand{\be}{\begin{equation}}
\newcommand{\ee}{\end{equation}}

\newcommand{\ba}{\begin{eqnarray}}
\newcommand{\ea}{\end{eqnarray}}

\begin{document}

\title{Power generation from a radiative thermal source \\ using  a large-area infrared rectenna}
\author{Joshua Shank}
\author{ Emil A. Kadlec}
\author{ Robert L. Jarecki}
\author{Andrew Starbuck}
\author{Stephen Howell}
\author{David W. Peters}
\author{Paul S. Davids$^{*}$}

\begin{abstract}
Electrical power generation from a moderate temperature thermal source by means of direct conversion of infrared radiation is  important and highly desirable for energy harvesting from waste heat and  micropower  applications.  Here, we demonstrate direct rectified power generation from an unbiased large-area nanoantenna-coupled tunnel diode rectifier, called a rectenna.  Using  a vacuum radiometric measurement technique with irradiation from a temperature-stabilized thermal source, a  generated power density of 8 nW/cm$^2$ is observed at a source temperature of 450C for the unbiased rectenna across an optimized load resistance.   The optimized load resistance for the peak power generation for each temperature coincides  with the  tunnel diode resistance at zero bias and corresponds to the impedance matching condition for a rectifying antenna.   
Current voltage measurements of a thermally illuminated large-area rectenna show current zero crossing shifts into the second quadrant indicating rectification.   Photon-assisted tunneling in the unbiased rectenna is modeled as the mechanism for the large short-circuit photocurrents observed where the photon energy serves as an effective bias across the tunnel junction.  The measured current and voltage across the load resistor as a function of the thermal source temperature represents direct current electrical  power generation.  

\end{abstract}

\affiliation{%
Sandia National Laboratories, Albuquerque, NM 87185, USA
}%
\date{\today}
\email{pdavids@sandia.gov}


\maketitle

\section{Introduction}

Conversion of electromagnetic radiation into electric current is the basis of electromagnetic detection\cite{sze2006physics,boreman}, wireless power transfer\cite{Hagerty,suh2002high,mcspadden1998design}, and solar energy harvesting\cite{sze2006physics,Yoshikawa:2017aa}. 
Photodetection and optical to electrical energy conversion  typically rely on photon-induced conductive currents in semiconductor device structures.  
The thermal infrared region of the spectrum lies at a cross-over point between optical photovoltaic conversion in photodiodes and direct power conversion through rectification in the radio-frequency region of the spectrum.   Infrared square-law detectors and energy conversion devices require small bandgap semiconductors matched to the infrared photon energy.  At room temperature, these small bandgap semiconductors are in thermal equilibrium with a large number of thermally generated photons.  Thermal source illumination from a higher temperature source in the bandwidth of the detector must generate more photons than the equilibrium photon gas\cite{landsberg1980thermodynamic}.   If photons from  the thermal source   do not out-number these thermally generated photons in the device then the signal is obscured by thermal noise and no useful work can be obtained from this device.   This is a consequence of the second law of thermodynamics.  For most infrared photodetection applications, these devices need to be cooled  to reduce the thermal background photons in order to produce a detected signal\cite{boreman}.  However, for this reason they are not good thermal energy harvesting devices for moderate temperature sources.   

Infrared direct detection  based on rectification of the displacement current in a tunnel diode can reduce the need for device cooling but requires the ability to respond on the time scale of the oscillation of the infrared radiation.  In the thermal infrared this corresponds to frequencies in the range of  20-50 THz, where carrier transport phenomena are typically too slow. Many research groups have examined ultrafast tunneling as a means of rectification and detection for optical frequencies\cite{ward2010optical, grover2010traveling, Sharma:2015aa, doi:10.1063/1.4995995}.  However, these optical rectennas typically are operated under high power laser illumination and applied DC bias,  and do not scale to large area readily.  By utilizing complimentary metal-oxide-silicon (CMOS) fabrication techniques,  uniform high-yielding distributed tunnel diode rectenna can be fabricated over full wafers with diameters of 300 mm.

\begin{figure*}[htb]
\centering \includegraphics[width=0.95\textwidth]{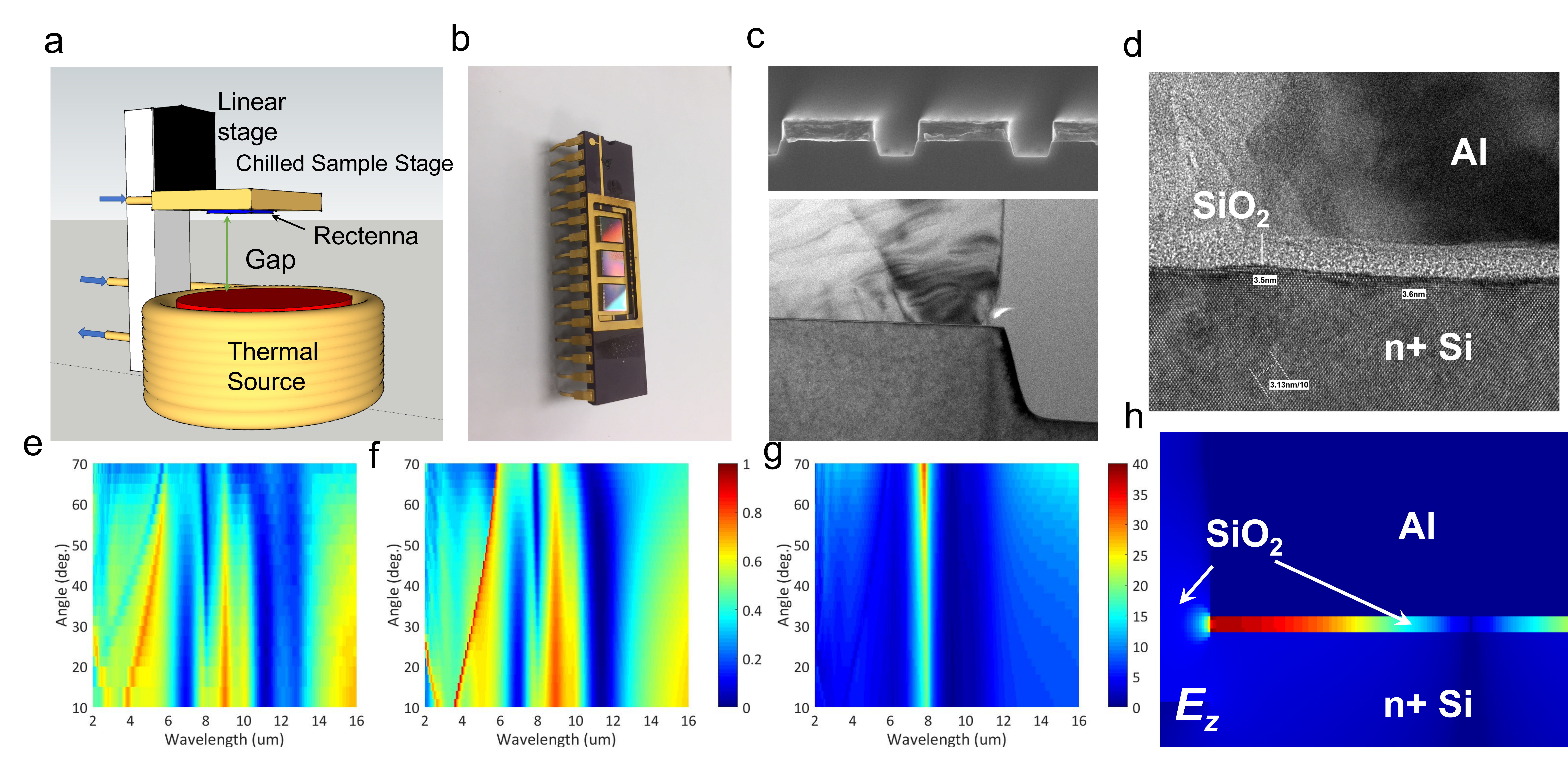}
\caption{{\bf Schematic of vacuum radiometric measurement for thermal source irradiation of a large-area infrared rectenna.}  (a) The schematic shows a water cooled shroud around a temperature controlled ceramic thermal heater that acts as a thermal source. A packaged rectenna  with active area 3 mm x 3 mm shown in (b) is attached to a chilled large Cu block that is on a linear stage that controls the source rectenna distance.  The temperature of the rectenna and the Cu block are monitored with attached thermo-couples. (c) The large-area rectenna is an integrated nanoantenna coupled metal-oxide-semiconductor tunnel diode that has been described previously\cite{davids2015infrared,kadlec_prapplied}. (d) High resolution TEM shows the tunnel diode  cross-section highlighting the oxide thickness ($\simeq$3-4nm).  The entire large-area nanoantenna coupled tunnel diode is encapsulated in thick oxide.  (e) Simulated (d) measured TM reflectance spectra from rectenna. (g) Simulated peak field concentration factor $\gamma =\text{E}_z^{gap}/\text{E}_0$ in rectenna. (h) Z-directed  transverse field concentration  in tunnel oxide at resonance.   }
\label{fig:1}
\end{figure*}

Tunneling diode rectifiers with high asymmetry are required to efficiently convert the small signal infrared electric field amplitude in an unbiased diode into a photocurrent.   Furthermore, the tunnel diode must be located near the concentrating infrared antenna to minimize  ohmic  loss associated with metals in the infrared.  This parasitic ohmic absorption leads to an increase in the temperature of the device and reduces the conversion efficiency.    In order to overcome these inefficiencies,  nanoantenna and metasurface design concepts  have been developed to create distributed tunnel diode rectifiers to resonantly concentrate the infrared electric field in the tunnel barrier for enhanced tunneling. 

Previously, we have demonstrated room temperature photoresponse of  large-area 1D and 2D nanoantenna-coupled tunnel diodes under coherent infrared illumination in biased and unbiased operation\cite{davids2015infrared,kadlec_prapplied}.  The infrared photoresponse was attributed to the large field enhancement in the spectral region near the epsilon near zero (ENZ) material resonance of the oxide tunnel barrier leading to enhanced tunneling rectification.   In this paper, we examine  a 1D rectenna  under thermal source illumination in a  vacuum radiometry experimental setup.   We demonstrate power generation across a load resistance and discuss photon-assisted tunneling in the unbiased device and optimization for thermal power supplies.

\section{Experiment}

\begin{figure*}[!htb]
\centering \includegraphics[width=0.8\textwidth]{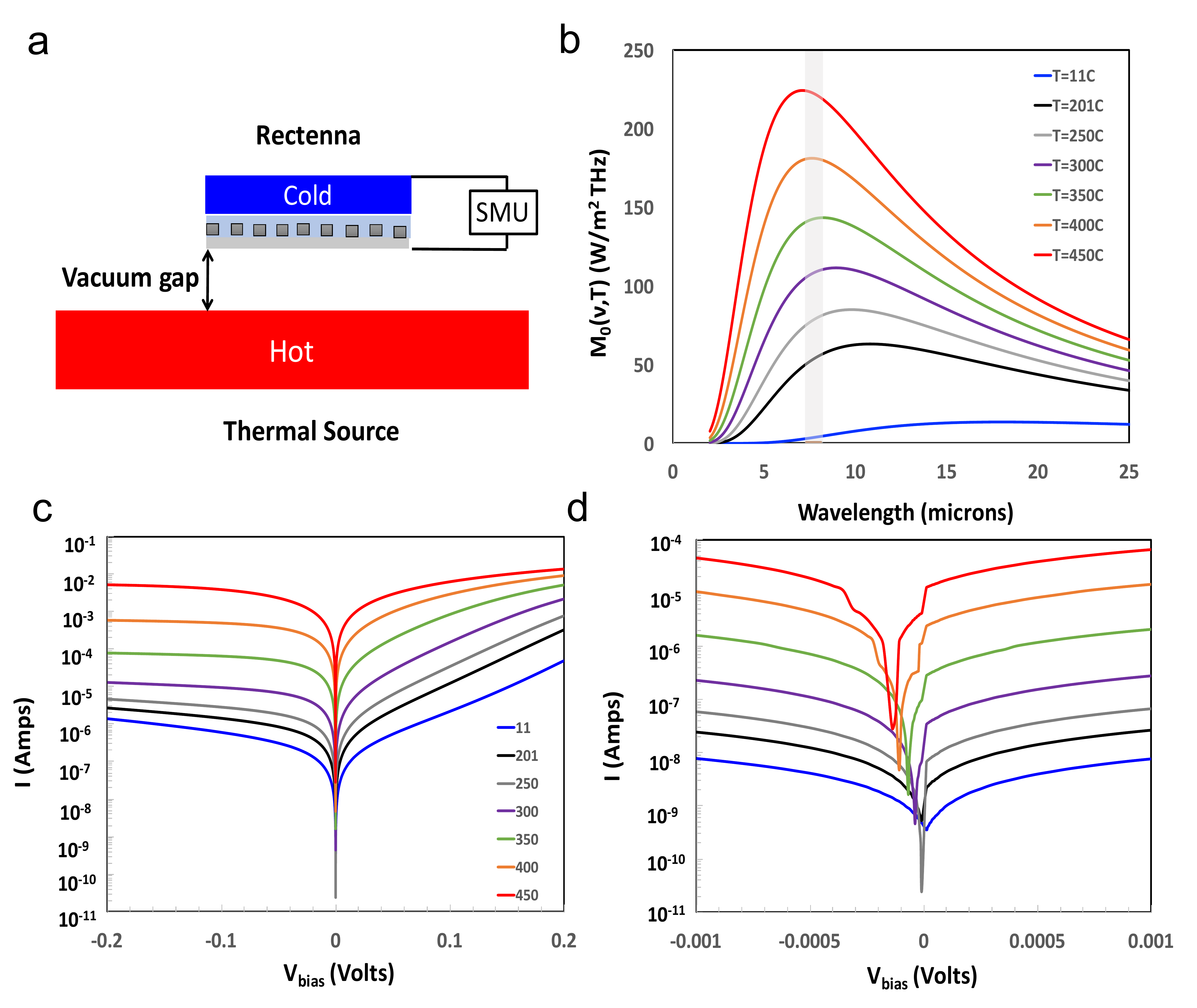}
\caption{{\bf Rectenna IV characteristics under thermal source irradiation.}  (a) The schematic of vacuum radiometric measurement of IV characteristics of rectenna under thermal source illumination. (b) Spectral blackbody exitance for blackbody source in the temperature range considered for the experiment. (c) IV sweeps of the rectenna at varying thermal source temperatures. (d) {\it High resolution} IV characteristics around zero bias showing the current minimum shifting into the second quadrant as the thermal source temperature increases.  Note: thermo-couple measurement of the rectenna does not show any significant heating of device or the Cu thermal sink.  }
\label{fig:2}
\end{figure*}

Vacuum radiometry is a technique used to measure photo-thermal response of thermo-photovoltaics and infrared rectennas under thermal illumination.  The experimental setup is shown schematically in figure \ref{fig:1}(a), where a ceramic heater with integrated temperature controller acts as a thermal source of radiation.   The heater is shrouded by a water-cooled coil and the emissivity of the heater is $\approx$ 90\% across the thermal infrared  band.   The packaged rectenna sample is mounted to a large water-cooled copper block and electrically connected to a ribbon cable for electrical characterization.   The cooled rectenna assembly is mounted to a linear stage which controls the distance between the thermal source and device under test.  (See figure \ref{fig:1}(b)) Thermo-couples are attached to sample to monitor temperature.   

The large-area rectenna  structure has been previously described and consists of an integrated, large-area metal grating antenna structure with a metal-oxide-semiconductor tunnel diode shown in figure \ref{fig:1}(c)-(d).  The grating is designed to effectively confine and enhance thermal infrared radiation into a 3-4 nm tunnel barrier.  Figure \ref{fig:1}(e) shows the simulated and measured (f)  transverse magnetic reflectance  for the rectenna.  A large absorption resonance is seen in the region near the oxide longitudinal optical phonon resonance at 8.1 $\mu$m.  The transverse field enhancement is shown in figure \ref{fig:1}(g) as a function of the wavelength and incident angle and the field enhancement profile near resonance at 7.3 $\mu$m, figure \ref{fig:1}(h),  is shown from simulation.  At normal incidence, the field enhancement factor is approximately 10-15$\times$ the incident field amplitude.  
The transverse field enhancement in the tunneling barrier is seen to increase for oblique incidence.  

The electrical characterization of the rectenna is  shown schematically in figure \ref{fig:2}(a) and illustrates the device under black-body illumination in vacuum.   The black-body spectral exitance incident on the rectenna is shown in  figure \ref{fig:2}(b) for the source temperature range considered. The spectral exitance is the illuminated power per unit area per unit bandwidth and can be compared to the Poynting vector flux irradiating the rectenna where the gray band indicates the absorption resonance on the rectenna.  Explicitly,  the spectral exitance,  $M^0_{\nu}$,  is given by 
\be
 M^0_{\nu}(T) = \frac{2\pi h \nu^3}{c^2} \frac{1}{\exp(\beta h \nu)-1}
\ee
where  $\beta = 1/k_B T$, $h$ is Planck's constant, $c$ is the speed of light, and $\nu$ is the photon frequency. From the spectal exitance,  the incident photon field amplitude,  $E_0$,  is derived from  its relationship with the Poynting vector flux incident onto the device.  As we have seen, the  transverse electric field amplitude in the barrier is enhanced relative to the incident field,  $E^{gap}_z = \gamma E_0$,  in the spectral region near the ENZ resonance, and is given by $ E^{gap}_z =\gamma (2Z_0 M^0_{\nu}(T)  d\nu)^{1/2} $,
where $Z_0$ is the impedance of free-space, d$\nu$ is the bandwidth, and $\gamma$ represents the transverse field enhancement factor in the tunnel barrier.  Figure \ref{fig:2}(c)-(d) show the current voltage characteristics for the thermally illuminated rectenna for source temperatures in the range of room temperature to 450C.  The IV data indicates large changes in the rectenna conductance as the source temperature is increased.  Furthermore, the IV near zero bias shows that the open-circuit voltage change shifts from zero  to negative voltage indicating rectification occurring in the 2nd quadrant.  The  large increase in tunneling conductance accompanies the change in diode asymmetry as the temperature increases and we observe large short-circuit currents at zero bias. 

\begin{figure*}
\centering \includegraphics[width=0.9\textwidth]{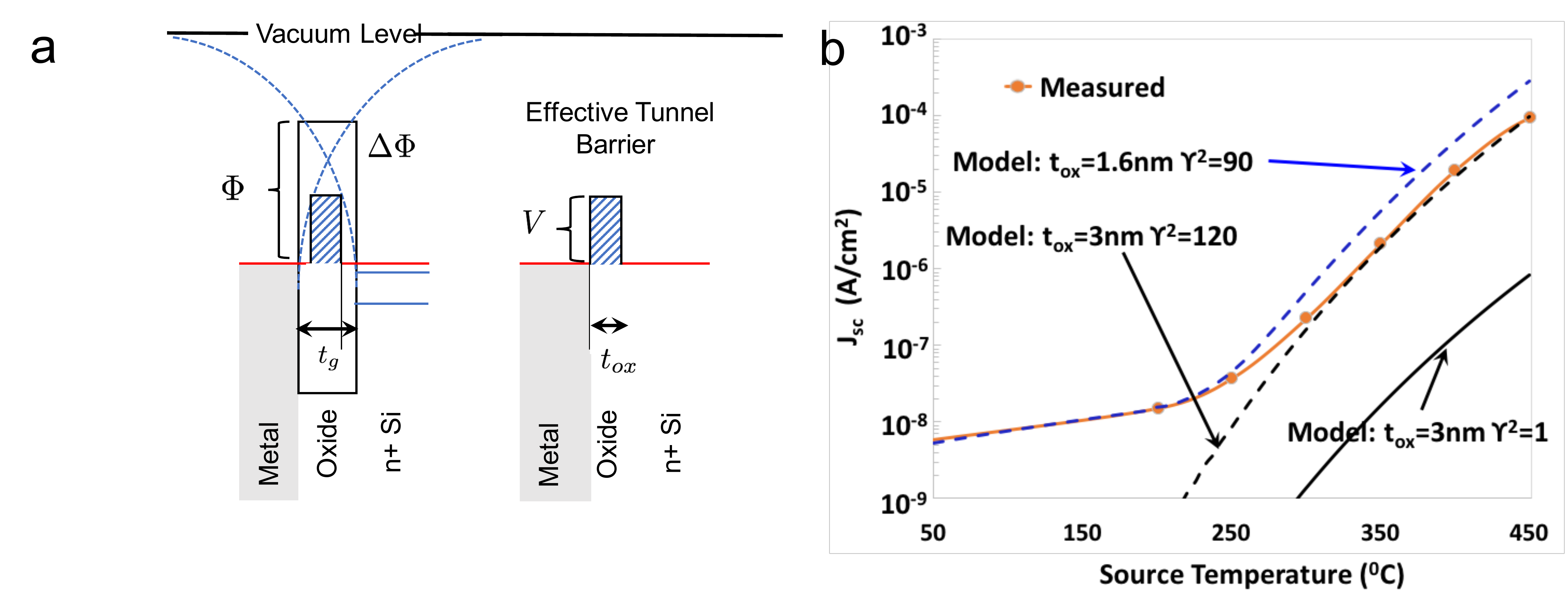}
\caption{{\bf Short-circuit rectenna current density as a function of source temperature.}  (a) Schematic band diagram of n-type MOS tunnel diode with Schottky image force lowering of tunnel barrier. The resulting effective barrier height and width are shown.  (b) Measured and modeled short-circuit current density for various effective barrier conditions.  $\Phi=3.0$ eV and $t_{gap}=4.0$nm for the Al/SiO$_2$/n+ Si MOS device. In all modeling, V=$1.65$ eV and the effective oxide thickness, $t_{ox}=3$ nm.  The longitudinal effective masses are  $m_l=m_e$, $m_r=0.19m_e$, and $m_{ox} = 0.5m_e$ where $m_e$ is the bare electron mass.  }
\label{fig:3}
\end{figure*}
   
Photon-assisted tunneling is a well established process is superconducting tunnel junctions\cite{Tien_Gordon_multiphoton,Bardeen_1962,Cohen_1962,exp_photon_tunneli} and has been observed in semiconductor superlattices\cite{PhysRevLett.75.4098}.  The short-circuit tunneling current can be  modeled by considering photon-assisted tunneling within the transfer Hamiltonian model for a uniform barrier tunneling under zero bias. The nanoantenna n-type MOS tunnel diode band diagram can be modeled as a uniform barrier that is reduced by Schottky image potential lowering yielding an effective barrier height and tunneling thickness.  Reductions in effective barrier height and thicknesses  for MOS tunnel diodes on degenerately doped Si substrates have been  extensively examined\cite{doi:10.1063/1.323415} and image potential lowering calculations can affect the computed tunnel currents\cite{Miskovsky1982}.  Figure \ref{fig:3}(a) shows the band-diagram for the n-type MOS tunnel diode illustrating the impact of image force lowering of the barrier on both sides of the device.  The bulk barrier height, $\Phi$=3.0 eV is lowered by 1.35 eV and the effective  tunneling distance at the Fermi level is reduced from the metallurgical thickness of approximately 4.0 nm to an effective tunnel thickness of 3.0 nm.  Photon-assisted tunneling is a dynamical tunneling effect, where the image charges are induced on the metal and semiconductor surfaces and dynamically screened at THz rates.  Therefore, the image potential calculation, uses the dynamic frequency-dependent permittivity in the resonance band of nanoantenna for the computation for the effective barrier parameters.  (See Appendix figure \ref{sfig:1}) Figure \ref{fig:3}(b) shows the measured short-circuit current as a function of source temperature.  An observable increase in the short-circuit current occurs near a source temperature of $T\simeq 250$C, where the peak of the black-body spectral exitance crosses the rectenna resonance. (see fig. \ref{fig:2}(b)) 

\begin{figure*}[!htb]
\centering \includegraphics[width=0.95\textwidth]{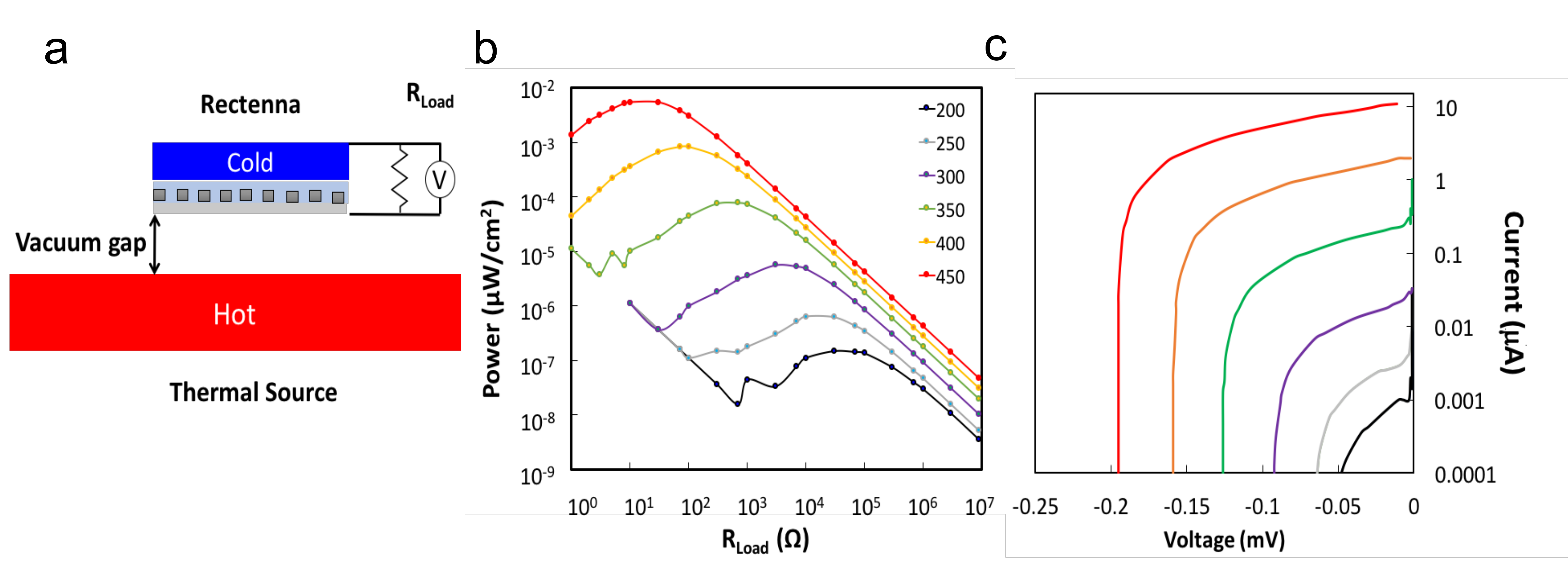}
\caption{{\bf Power generation from a thermal source in an external load resistor.}  (a) Schematic of vacuum radiometry experiment for power generation.(b) Power generation as a function of the load resistance for various source temperatures.  (c) Load diagram in second quadrant illustrating the current and voltage in the load resistor as a function of the source temperature.   }
\label{fig:4}
\end{figure*}

The expression for the single photon contribution to the short-circuit current density is \cite{davids_shank}
\begin{widetext}
\be
J^{(1)} = AT^2 \left( \frac{m_r m_l}{m^2_{ox}} \right)  2 Z_0 \int^{\nu_{max}}_{\nu_{min}} d\nu  (1-R^2)\gamma^2  \left(\frac{e t_{ox}}{ h \nu} \right )^2 M^0_{\nu}(T)  ~{\cal T}(\nu) ,
\label{eq:current_1}
\ee
where $A =120 \text{ Amps}/ \text{cm}^2 \text{K}^2$ is  Richardson's constant, and 
\end{widetext}
\be
{\cal T} (\nu) =  \beta \int dE_l~t(E_l,E_l+  h \nu ) \ln \left[  \frac{ 1+e^{-\beta E_l} }{1+ e^{-\beta(E_l  + h \nu)}   }  \right]
\label{eq:current_2}
\ee
is the integrated barrier transmission,  and $h \nu$  is the photon energy that acts as an effective bias in the supply function.  The transfer Hamiltonian barrier transmittance is 
\be
t(E_l,E_r) =e^{-(k+q)t_{ox}}~ \text{sinhc}^2\left(\frac{(k-q)t_{ox}}{2} \right),
\ee
where the first term has the form of the standard tunneling exponential and the second term is the square hyperbolic sinc function. The evanescent wavevectors in the barrier region are  $k =(2m_l (V -E_l)/\hbar^2)^{1/2}$ and $q =(2m_r (V -E_r)/\hbar^2)^{1/2}$ as defined in caption.  The modeled zero bias current density is shown in figure \ref{fig:3}(b) with various field concentration factors and effective oxide thicknesses.  The modeled results accurately predicts the temperature transition and the short-circuit current density above the transition temperature for realistic effective barrier height,thicknesses, and computed field enhancement factors for normal incidence.  

Finally, we come to the ultimate goal of electrical power generation from radiated heat from a thermal source.   Figure \ref{fig:4}(a) shows the configuration for the power generation measurement.  An external variable resistive load is placed across the thermally illuminated large rectenna and a voltage is recorded.  In the experiment, the thermal source-rectenna  vacuum gap of $\simeq$2-3 mm is fixed for all measurements, and the external resistor board is outside the vacuum at room temperature.   Thus the measurement is in the far-field of the thermal source and due to the large size difference between the heater and the rectenna, the view factor is approximately unity out to large angles of incidence. The measured voltage and current is recorded and the electrical power delivered to the load resistor computed.  Figure \ref{fig:4}(b) shows the electrical power generated as the load resistance is varied for a set of thermal source operating conditions.   The peak power is seen to shift to lower resistance values as the source temperature is increased.   The peak load resistance matches and tracks the tunnel diode resistance at zero bias indicative of direct rectification impedance matching requirements.  Electrical rectification from an AC signal requires that the diode impedance match the load impedance for maximum power transmission to the load.   The current voltage load diagram is shown in figure \ref{fig:4}(c).  The power generation is clearly in the second quadrant and photo-generated currents of 10 $\mu$A and voltages of tenths of a mV are obtained at the higher source temperatures.   

\section{Conclusions}
In summary, we have unambiguously generated electrical power from a radiative thermal source at moderate temperatures.   The measured currents and voltages across a load impedance are indicative of direct rectification of infrared radiation in our unbiased large-area nanoantenna-coupled tunnel diode.   The peak power load resistance is seen to precisely match the tunnel diode resistance at zero volts, consistent with antenna rectifier impedance matching conditions.  A model of photon-assisted tunneling through an effective tunnel barrier has been proposed and quantitatively matches the short-circuit current over three decades.  The model includes realistic field concentration in the tunnel barrier as observed under finite-element modeling of the electromagnetic response of the device.  This high field in the tunnel barrier gives rise to the large short-circuit currents and can be tuned through design of the metasurface resonance\cite{liu2010infrared,alu2007epsilon,Pendry2004b} and the material optical phonon modes\cite{PhysRevApplied.4.044011}.  While the power generation seems modest, the large-area fill-factor can compensate for low efficiency conversion and  improvements in the device design and improvements  in the device performance can be readily implemented within a standard CMOS process.  The result would be a new micropower supply that could convert radiated waste heat into useable electric power for wide ranging applications.


\acknowledgements{
Funding for this work was provided by Sandia's Laboratory Directed Research and Development
(LDRD) program. Sandia National Laboratories  is a multi-mission laboratory managed and operated
by National Technology and Engineering Solutions of Sandia, a wholly owned subsidiary of Honeywell International Inc., for the United States Department
of Energy's National Nuclear Security Administration under contract DE-NA0003525.
}

%


\appendix
\begin{figure*}[htb]
\centering \includegraphics[width=0.95\textwidth]{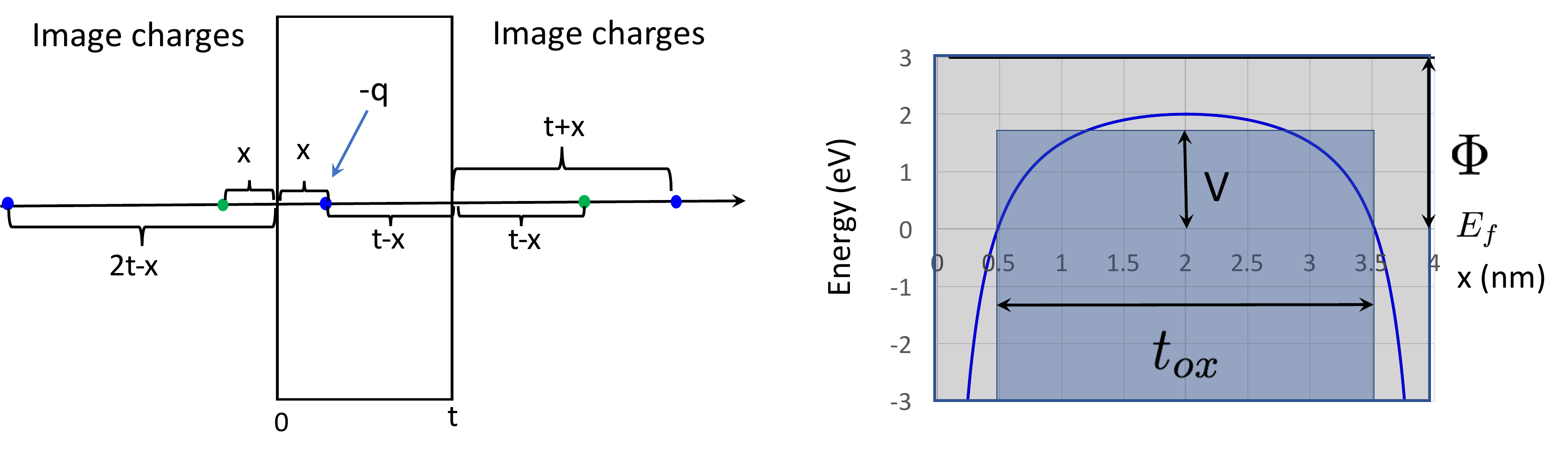}
\caption{{\bf Image potential calculation and effective barrier height and width}    (a) Method of image solution for a charge located at position $x$ from the left metal contact after 2nd stage for the metal-vacuum-metal structure. (b) Image potential lowered barrier and effective rectangular barrier shown. The bulk Al/SiO$_2$ barrier height is 3.0 eV and the gap thickness is 4 nm.  The effective uniform barrier height is 1.65 eV and the thickness is 3 nm.  }
\label{sfig:1}
\end{figure*}

\section{ Image potential barrier lowering}
\label{App:1}
The tunneling barrier shape, height $V$ ,and width $t_{gap}$ are modified by the metal image potential in the metal-insulator-degenerately doped Si tunnel diode\cite{Miskovsky1982}.  To solve for the electric potential of an electron in the tunnel barrier, we  use the method of images to compute the potential, $\phi$, and the electric energy $W=q\phi/2$ at the position of a test charge in the dielectric region of a MIM diode.   Figure \ref{sfig:1}(a) shows the configuration for a charge $-q$ at position $x$.   The images charges are placed to make the two surfaces equipotential surfaces and this process can be iterated.   The resulting image potential energy is 
\be
W = -\frac{q^2}{8\pi \epsilon_0 \epsilon} \left\{ \sum_{n=1}^{\infty} \left(\frac{nt}{(nt)^2 -x^2} -\frac{1}{nt} \right)   +\frac{1}{2x}\right\},
\ee
and the effective barrier $\Phi_{e}(x) = \Phi + W(x)$.   The oxide barrier height is set at the bulk barrier for Al on SiO$_2$,  $\Phi =3.0$eV. The image potential lowered barrier, $\Phi_{e}(x)$, is shown in figure \ref{sfig:1}(b)  and can be converted into an effective rectangular barrier by integration. We obtain
\be
V\cdot t_{ox} = \int_{x0}^{x_1} dx \Phi_{e}(x),
\ee
where $\Phi_{e}(x_0)=0$ and $\Phi_{e}(x_1)=0$ are the intersections of the barrier at $E_f=0$.   Typically, the relative permeability for the oxide is used in the calculation, but in the rectenna device the tunneling is dynamic.  The enhanced electric field in the barrier is inducing image charges at 30 THz, and it therefore makes sense to use the dispersive infrared permittivity in the calculation.    In the infrared resonance band, $\epsilon $ varies from 1 $\rightarrow$ 0, so for the calculation of the effective uniform barrier, we use $\epsilon=1/2$.  From the above discussion, the  effective uniform barrier height $V=1.65$eV and the effective oxide thickness, $t_{ox}=3$nm as described in the manuscript.

\section{Impedance match}

\begin{figure}[htb]
\centering \includegraphics[width=0.5\textwidth]{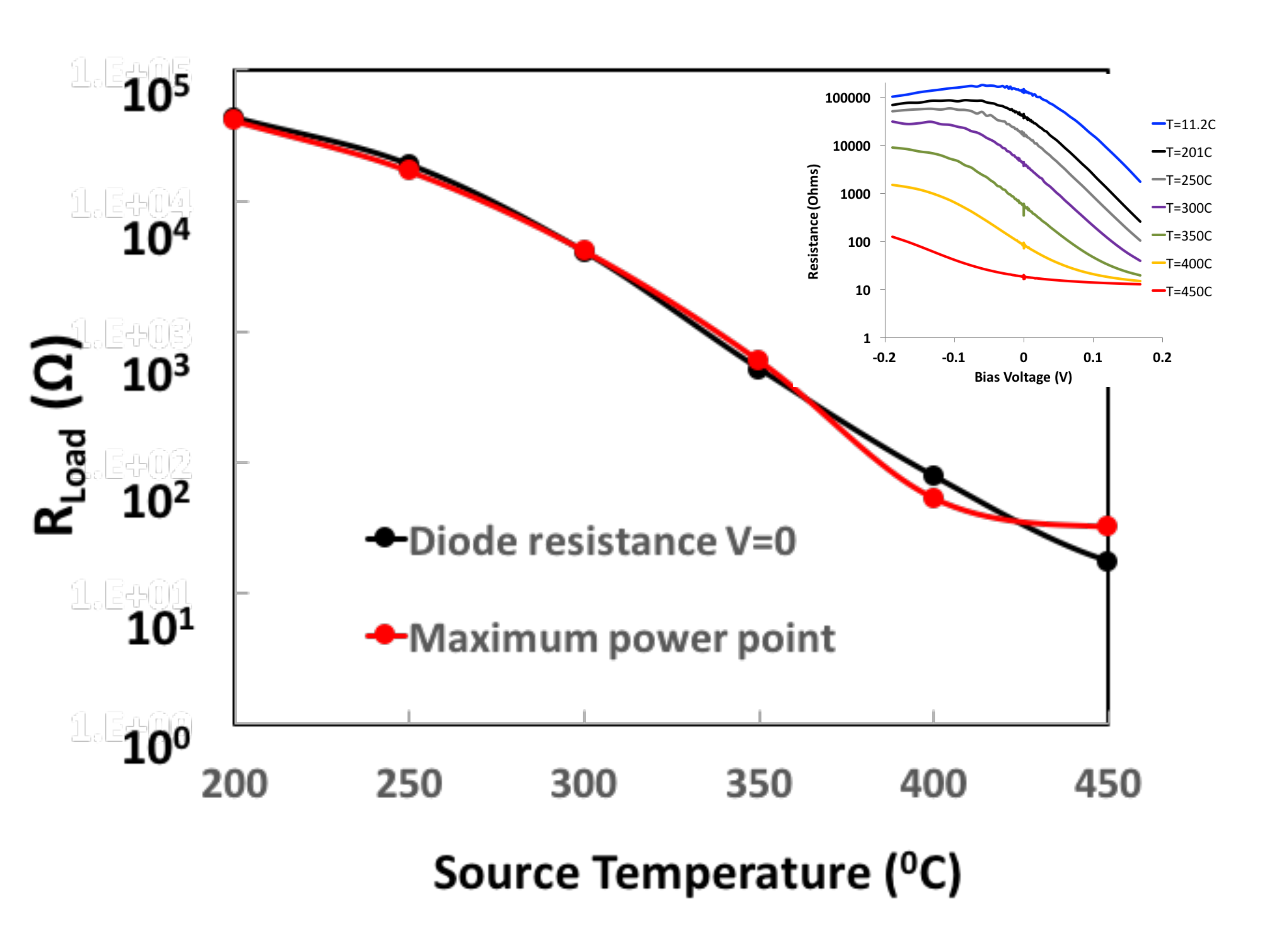}
\caption{{\bf Impedance matching and diode resistance versus source temperature}  Load resistance (left) at peak power generation as a function of the source temperature.   The diode resistance (inset) as a function of applied voltage under illumination from a thermal source.  The diode resistance at zero bias for each source temperature is also plotted in the left panel.}   
\label{sfig:2}
\end{figure}
The conditions of impedance matching are well developed in rectified detection and power sources.  Antenna matching networks are used to reduce reflection from the antenna to maximize power to the rectifying diode, and a second matched network is used to maximize power to the load impedance.   In the rectenna, the diode rectifier is distributed and integrated into the antenna surface.  The rectenna is optimized to have low reflectance in the region where we have high field enhancement in the tunnel barrier.   The condition for maximal power generation occurs when the load resistance matches the tunnel diode resistance at zero bias.   Figure \ref{sfig:2} shows that the rectenna satisfies the impedance matching condition.  Typically, you achieve peak power delivered to the load when the diode impedance matches the load impedance.  The peak power load resistance matches over many decades of resistance the tunnel diode resistance at V=0 as a function of the source temperature.  This clearly indicates that the power generation arises from rectification and not from bolometric or internal thermoelectric effects.

\section{Figure of merit}

\begin{figure}[htb]
\centering \includegraphics[width=0.45\textwidth]{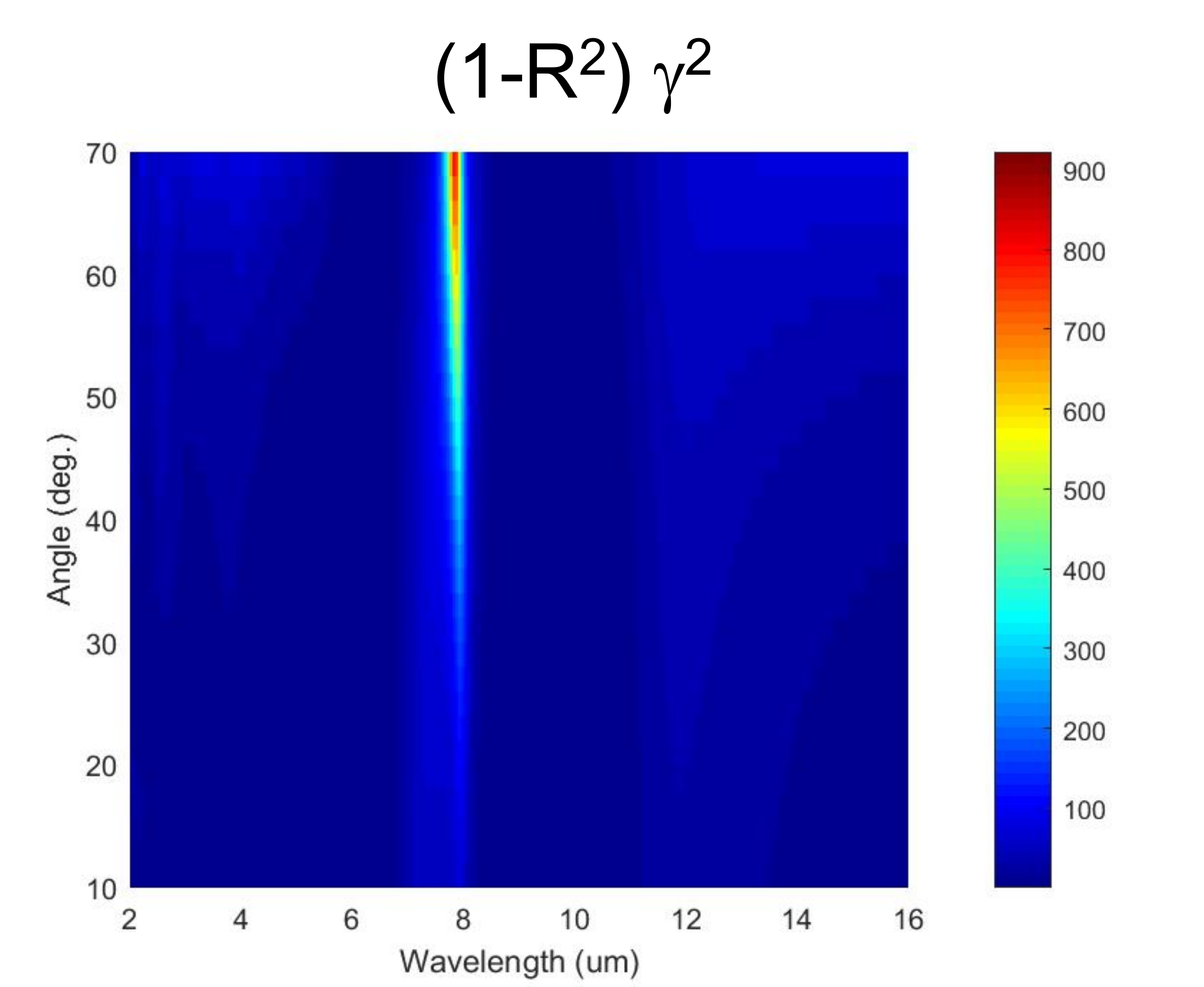}
\caption{{\bf Figure of merit for resonant field enhancement}  The computed figure of merit as a function of the incident angle and wavelength for the grating nanoantenna structure outlined in the paper.   }
\label{sfig:3}
\end{figure}

The figure of merit, $F(\theta, \lambda) = (1-R^2)\gamma^2$, represents the field enhancement in the tunnel gap weighted by the absorption of the nanoantenna and is in the integrand for the photon-assisted tunnel current expression Eq.2.  In the computed short-circuit current model, we use fixed normal incidence values for the figure of merit. Ideally, we want high absorption in the nanoantenna structure  and high field concentration in the tunnel gap.   These are competing requirements since the field concentration in the tunnel gap occurs very close to the ENZ LO phonon mode of the oxide. It is difficult to obtain zero reflection from the oxide over-coated structure near the ENZ point.   Figure \ref{sfig:3} shows the computed figure of merit for the nanoantenna structure of the paper as a function of angle and wavelength.  It is instructive to note that at large angles we get the largest figure of merit near the ENZ resonance of 8.1 $\mu$m for the oxide.  This field enhancement is sizeable for oblique incidence near the well known Berreman guided mode resonance\cite{berreman1963infrared}.   This implies that power generation at oblique incidence will be significant and  we can optimize the view factor to improve the conversion efficiency.

\end{document}